\documentclass[%
 reprint,
superscriptaddress,
 amsmath,amssymb,
 aps,
]{revtex4-2}

\usepackage{graphicx}% Include figure files
\usepackage{dcolumn}% Align table columns on decimal point
\usepackage{bm}% bold math
\usepackage{hyperref}% add hypertext capabilities
\hypersetup{colorlinks=true, citecolor=blue, urlcolor=blue, linkcolor=blue}
\usepackage{overpic}

\usepackage{xr-hyper}
\usepackage{xcolor}
\usepackage{soul}
\usepackage[utf8]{inputenc}
\usepackage{siunitx}
\usepackage{ragged2e}
\usepackage{booktabs}
\usepackage[justification=justified,singlelinecheck=false,compatibility=false]{caption}
\usepackage{subcaption}

\DeclareSIUnit{\electronvolt}{eV}

\begin{document}

\author{Ludwig Hedin}
\email{ludwig.hedin@liu.se}
\affiliation{% 
Theoretical Physics Division, \\
Department of Physics, Chemistry and Biology (IFM),
Link\"{o}ping University, SE-581 83, Link\"{o}ping, Sweden
}%

\author{Johan Klarbring}
\email{johan.klarbring@liu.se}
\affiliation{% 
Theoretical Physics Division, \\
Department of Physics, Chemistry and Biology (IFM),
Link\"{o}ping University, SE-581 83, Link\"{o}ping, Sweden
}%
\title{ Barocaloric phase transformation from data efficient fine-tuning of machine learned interatomic potentials}% Force line breaks with \\

\date{\today}% It is always \today, today,

\begin{abstract}
Solid-state cooling based on the barocaloric (BC) effect has emerged as promising environmentally friendly prospective alternative to conventional vapor-compression refrigeration. The search for suitable BC materials relies on efficient atomistic simulations of their phase behavior. Machine-learned interatomic potentials (MLIPs) enable such simulations at near density functional theory (DFT) accuracy, but generating the required DFT training data remains computationally demanding, which motivates development of strategies that reduce the amount of data needed to train accurate models. In this work, we use a prototypical barocaloric material, ammonium sulfate, as a model system and investigate how small a training set can be while still reproducing the temperature-driven structural phase transformation that underlies its barocaloric response. We train a series of MLIPs based on the MACE architecture using three strategies: training from scratch, and naive- and multihead replay fine-tuning of the MACE-MPA-0 foundation model. These strategies are evaluated on their ability to reproduce the phase transformation of ammonium sulfate in molecular dynamics simulations across a range of training-set sizes. We find that, while the MACE-MPA-0 foundation model itself fails to reproduce the transformation, and models trained from scratch break down for small datasets, fine-tuned models reproduce the transformation using as few as 5 to 10 60-atom DFT configurations. Both fine-tuning protocols yield similarly accurate results for ammonium sulfate, but we also find some indications that multihead replay fine-tuning is more robust on configurations outside the fine-tuning domain. Exploiting this data efficiency, we further show that models can be trained on small datasets and the hybrid-DFT level, and that some form of inclusion of dispersion correction is necessary to describe the phase behavior correctly.
\end{abstract}

\maketitle

%\tableofcontents

\section{\label{sec:intro} Introduction}
The hydrofluorocarbon (HFC) refrigerants used in conventional vapor-compression cooling are potent greenhouse gases, with global warming potentials thousands of times greater than that of $\mathrm{CO_2}$~\cite{UN_cooling_report2020}, and with refrigeration consuming a large and growing share of global energy~\cite{IIR_report_2025}, more sustainable alternatives are needed. Solid-state cooling based on caloric materials is one such alternative, exploiting external field driven phase transformations that produce substantial temperature changes under adiabatic conditions. While magnetocaloric compounds have historically been the most extensively studied~\cite{Cazorla_mechanocaloric, Cirillo_BC_cooling, Giant_MC_effects}, barocaloric (BC) materials, where the transformation is driven by hydrostatic pressure, have recently attracted strong interest. Indeed, large barocaloric effects have recently been demonstrated in a range of material classes, including, for example, colossal effects in plastic crystals of neopentyl glycol (NPG)~\cite{neopentyl, Li2019}.

The barocaloric response of a material is governed by a phase transformation, and identifying promising candidates therefore requires characterizing this transformation. Atomistic simulations have become an indispensable tool for this task, allowing the relevant phase behavior to be characterized without recourse to experiment.  Classical molecular dynamics (MD) enables efficient large-scale simulations but is often hindered by the limited accuracy of the underlying interatomic potential. In contrast, density functional theory (DFT) provides accurate calculations, yet remains limited to relatively small systems of \(\sim10^3\) atoms and short time scales, due to its inherent computational complexity~\cite{Zuo_MLIP_performance_cost, Pan_DFT_scaling}. Machine learned interatomic potentials (MLIPs) promise to circumvent these issues~\cite{Adv_chal_MLIP, MLIP_materials_science, MLIP_practical_guide, MLIP_roadmap}. Indeed, modern MLIP architectures, when trained on well-constructed DFT-datasets of energies, forces and stress, retain near DFT-accuracy, while reducing computational cost by orders-of-magnitude~\cite{Zuo_MLIP_performance_cost}. 

However, despite the success of MLIPs, there still remains significant ambiguity regarding how to optimize training strategies. Moreover, obtaining high-quality training data from DFT computations demands substantial computational resources. Consequently, it becomes of central importance to employ training strategies aimed at reducing the amount of required training data without sacrificing accuracy.

So called foundation, or universal, MLIPs, trained on large chemically and structurally diverse datasets have gained prominence over the last few years~\cite{yuan2026foundation}. These models offer qualitative accuracy and stability in MD simulations over a wide range of chemistries and structures, but do not offer the same quantitative accuracy for specific material systems as custom-trained MLIPs on targeted datasets~\cite{MACE_foundation_model}.

Fine-tuning these pre-trained foundation models, rather than training an MLIP from scratch, has emerged as a promising route to data efficient training, and recent work has established that it improves accuracy across a range of MLIP architectures and target properties~\cite{Hnseroth2026,Kaur2025,Grandel2026,Radova2025,Lu2025}.

In this work, we build on these insights and use the prototypical BC material ammonium sulfate~\cite{ammonium_sulfate2015} as a model system. We ask specifically, how small can we make the training set, while still faithfully reproducing the structural phase transformation which underlies the functionality of barocaloric materials? We train a series of MLIPs based on the MACE architecture~\cite{MACE} using  different strategies and dataset sizes. Specifically, we investigate three training strategies: training MACE models from scratch, as well as fine-tuning a MACE foundation model using two distinct approaches: multihead replay fine-tuning and naive fine-tuning. These models are then evaluated on their ability to  reproduce the temperature-driven first-order phase transformation in ammonium sulfate from a series of MD simulations.  

We find that fine-tuned models reproduce the qualitative behavior of the transformation down to training sets of just five 60-atom DFT configurations. We further exploit this data efficiency and showcase that models can be efficiently constructed at beyond semi-local DFT accuracy by training models on small datasets of hybrid PBE0+D3 calculations. We find that some level of dispersion correction is essential in describing the phase behavior correctly, consistent with the transformation being driven by a reorganization of the hydrogen-bond network.

\section{\label{sec:methods} Methods}

\subsection{Ammonium sulfate phases}
\begin{figure}[h]
    \centering
    \begin{subfigure}{0.48\textwidth}
        \centering
        \begin{overpic}[width=\linewidth]{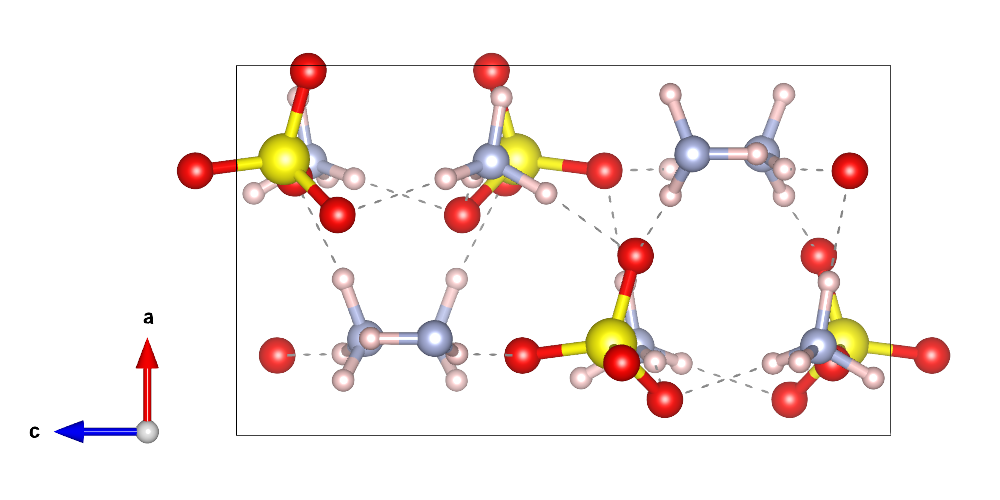}
            \put(3,40){(a)}
        \end{overpic}
    \end{subfigure}
    \begin{subfigure}{0.48\textwidth}
        \centering
        \begin{overpic}[width=\linewidth]{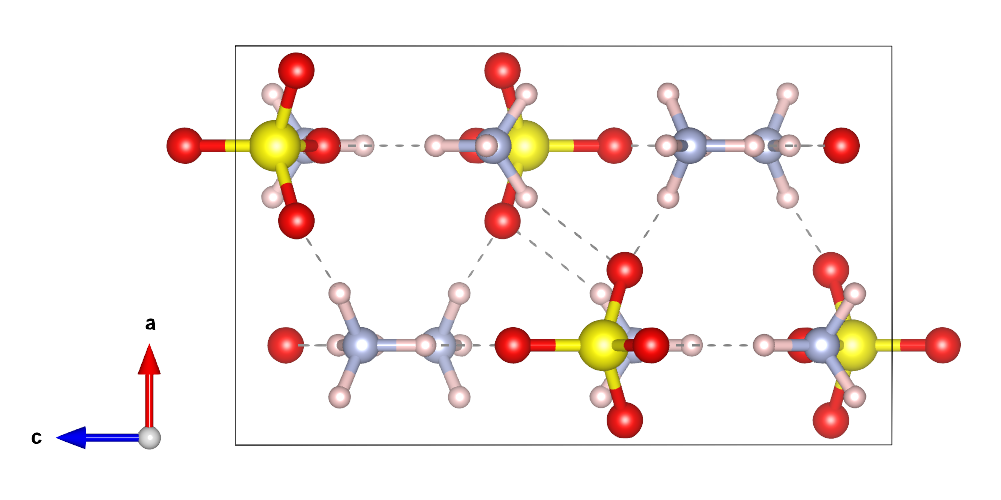}
            \put(3,40){(b)}
        \end{overpic}
    \end{subfigure}
    \caption{\justifying\justifying\justifying Unit cell of ammonium sulfate in (a) the low-temperature $Pna2_1$ phase, and (b) the high-temperature $Pnam$ phase. H, N, O and S atoms are shown in white, blue, red, and yellow, respectively. }
    \label{fig:as_phases}
\end{figure}

The barocaloric effects in ammonium sulfate, discovered by Lloveras \textit{et al.}~\cite{ammonium_sulfate2015} in 2015, arise from a pressure-induced structural phase transformation at $\sim$~\SI{219}{\kelvin}. In the high-temperature regime, ammonium sulfate stabilizes in a centrosymmetric orthorhombic structure ($Pnam$), while it adopts an orthorhombic polar structure ($Pna2_1$) when cooled below the phase transformation temperature, see Fig.~\ref{fig:as_phases}. As the barocaloric effects in ammonium sulfate have been extensively studied~\cite{ammonium_sulfate2015, ammonium_sulfate2021, ammonium_sulfate2022}, it serves as an ideal model system for this study.

\subsection{\label{sec:datasets}Dataset generation}
We initially generated two large and diverse datasets of relevant configurations. The first set contains 360-atom structures generated by running MD using the VASP machine learning force field (MLFF) implementation~\cite{Jinnouchi2019,Jinnouchi2019b,Jinnouchi2020}, where relevant snapshots are picked out and added to the dataset based on an on-the-fly evaluation of the model error. Separate runs were performed for a range of temperatures and pressures in both the low-temperature (LT) and high-temperature (HT) phases. This dataset contains 1385 structures, out of which 139 structures were set aside to construct a hold-out test set. A second set with 60-atom structures were obtained by running MD with the MACE-MPA-0~\cite{mace_foundation_models_git} foundation model at a set of temperatures and pressures, initialized in both the LT and HT phases, and randomly picking out a set of temporally well-separated structures. This dataset contains 2403 structures. Further details on the generation of these datasets are given in the Supplementary Material (SI).

Based on the 60-atom dataset we generated a series of subsets containing an increasing number of structures. This was done using the \textit{DImensionality-Reduced Encoded Clusters with sTratified} (DIRECT) sampling workflow proposed in Ref.~\cite{DIRECT_sampling}. In our case the configuration space is represented through the invariant features of the MACE-MPA-0 foundation model~\cite{mace_foundation_models_git}. Principal component analysis (PCA) is then applied to reduce the dimensionality of the feature space, followed by clustering to group configurations with similar features. Finally, 1 data point from each cluster is selected based on the euclidean distance to the center of the respective cluster.

\subsection{DFT calculations}
All DFT-calculations were performed in the projector augmented wave (PAW)~\cite{PAW} formalism using the VASP-code~\cite{vasp_96_1, vasp_96_2, vasp_99}. We performed DFT calculations on the full datasets using the r$^{2}$SCAN meta-GGA functional~\cite{r2SCAN}. We further did hybrid DFT calculations using the PBE0~\cite{PBE0} functional, with and without D3~\cite{D3} van der Waals corrections, on select subsets. 

For the r$^{2}$SCAN calculations, a plane-wave energy cutoff of \SI{900}{\electronvolt} was used. Brillouin zone sampling was restricted to the $\Gamma$-point for the 360-atom cells, for the 60-atom cells k-points were generated by means of \texttt{KSPACING}=0.5, while for the out-of-domain test-set \texttt{KSPACING}=0.44 was used. We used a 50 meV Gaussian smearing and electronic self-consistency was achieved with an energy convergence criterion of \SI{1e-6}{\electronvolt}. We used the VASP recommended set of standard PAW potentials, with valence configurations H ($1s^1$), N ($2s^2 2p^3$), O ($2s^2 2p^4$), and S ($3s^2 3p^4$).

\texttt{PRECFOCK=Normal} was used for the PBE0 and PBE0+D3 calculations, and all other parameters were consistent with the r$^{2}$SCAN calculations. 

Energies for single, isolated atoms were calculated using spin-polarized r$^{2}$SCAN and PBE0. We used large and slightly deformed supercells to avoid artificial high symmetry solutions.  

\begin{figure*}
    \includegraphics[width=\linewidth]{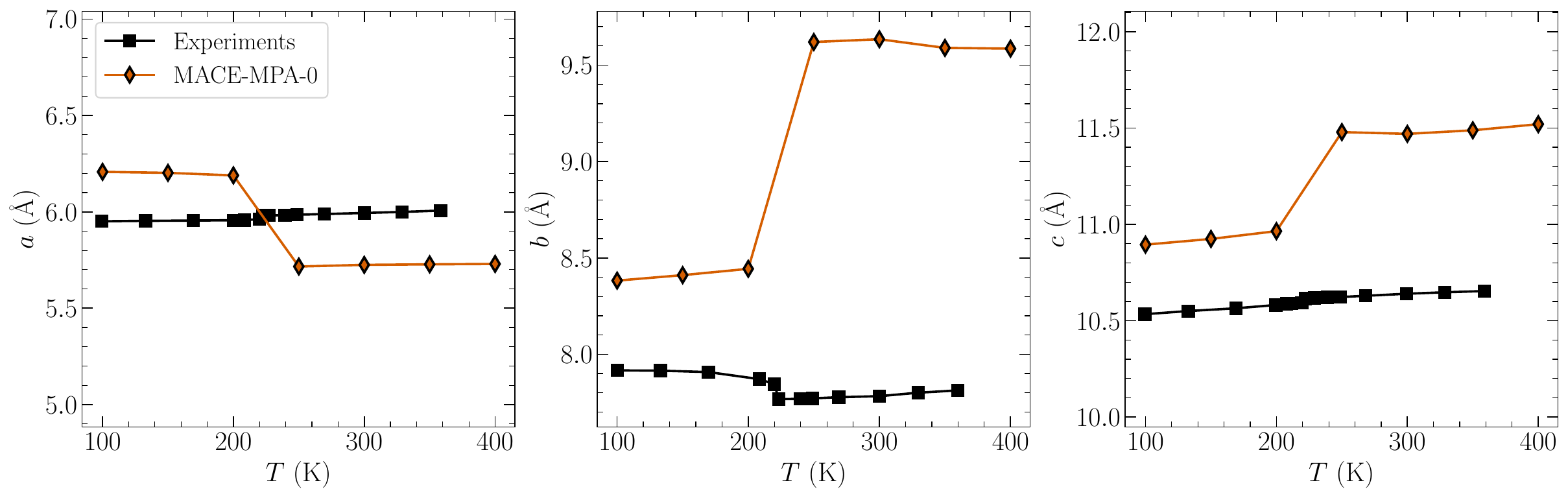}
    \caption{\justifying Temperature dependence of the lattice parameters $a$, $b$, and $c$ in ammonium sulfate, obtained from MD simulations using the MACE-MPA-0 foundation model. Experimental values of the lattice parameters from Ref.~\cite{ammonium_sulfate2015}.}
    \label{fig:mace-mpa}
\end{figure*}
 
\subsection{MACE models, training and fine-tuning}
We use MACE~\cite{MACE} version 0.3.12 for constructing MLIPs throughout this work. The MACE architecture is an equivariant message-passing neural network~\cite{MPNN}. Its message construction is based on the atomic cluster expansion (ACE)~\cite{ACE}, enabling high body-order equivariant~\cite{E3equivariance} features. The invariant features are then mapped to an invariant energy, resulting in efficient and accurate MLIPs.

Two reference models, using the full 360- and 60-atom datasets described above, respectively, were trained from scratch, i.e. starting from randomly initialized model weights.

We further constructed a series of MLIPs based on fine-tuning the MACE-MPA-0~\cite{mace_foundation_models_git} foundation model. The MACE-MPA-0 model is trained on a combined dataset including the MPtrj~\cite{MPtrj} and sAlex~\cite{sAlex,Omat24} datasets. 
Two separate fine-tuning protocols were employed: \\
\emph{Naive fine-tuning} (NFT), where one starts from the foundation-model weights and continues training on a custom dataset, and \\
\emph{Multihead replay fine-tuning} (MHFT), where one not only fine-tunes the model on the custom dataset, but also on a so-called replay dataset~\cite{MACE_foundation_model}. The replay dataset is selected to optimize the robustness of the fine-tuned model by sampling configurations from the pre-training set that contain elements appearing in the fine-tuning dataset. The model is then trained using both datasets simultaneously, sharing the model descriptor weights while constructing separate readout functions, or "heads"~\cite{MACE_foundation_model}. The MHFT protocol is expected to aid in preventing so-called "catastrophic forgetting" of the pre-trained model descriptors~\cite{MACE_foundation_model}.

As our aim is to compare the fine-tuned models to the from-scratch trained reference models and since the hyperparameters of the fine-tuned model are constrained to those originally used to train the foundation model, we used hyperparameters consistent with those of MACE-MPA-0 for all our models. Specifically, this implies a model with two message passing layers, correlation order 3, number of channels $k=128$, angular resolution $l_{max}=3$, equivariance order $L_{max}=1$ and cutoff radius $r_{max} = \SI{6.0}{\angstrom}$. We fine-tuned models with both training strategies on the subsets of our large dataset, constructed as described in Sec. \ref{sec:datasets}, with number of samples ranging from 5 to 500.

In training, an 80-20 split between training and validation data was consistently used for each training strategy. We used a hold-out test set that contains 10\% of the 360-atom dataset (139 structures) to evaluate all models trained on any of the $\mathrm{r^2SCAN}$ datasets. For MHFT, a replay training set containing 50000 structures from MPTrj was used, these contained 339 structures with any combination of the elements H, S, N and O and a random selection making up the rest. 

When training models from scratch, we set the maximum number of epochs to 500. Default values for the force, energy, and stress weights, which are 100, 1, and 1, respectively, were used. In addition, a loss scheduler was activated for the final 20\% of the training, in which the force, energy, and stress weights were changed to 100, 1000, and 10, respectively. A learning rate of 0.01 was used for the first part of the training, while it was set to 0.001 for the second stage of training. The batch size was set to 4.

In NFT, the maximum number of epochs was set to 500 and a learning rate of 0.0005 was used. We used force, energy, and stress weights of 10, 10, and 1 respectively. The batch size was set to 2. In MHFT, we set the number of epochs to 20 and the learning rate to 0.0001. The force, energy, and stress weights were set to 10, 10, and 1 respectively, and the batch size was set to 8.

\subsection{Molecular Dynamics}
All MD simulations were performed using i-PI~\cite{i-PI}, where the MACE models are interfaced using an ASE client. The simulations were carried out in the NpT ensemble using a Martyna-Tuckerman-Tobias-Klein barostat~\cite{barostat}. The barostat had a time constant of \SI{100}{\femto\second}, and both the ionic and barostat degrees of freedom were coupled to Langevin thermostats~\cite{langevin} with relaxation times of \SI{100}{\femto\second}.
We used a time step of \SI{0.5}{\femto\second}. All MD simulations were started from the LT phase at all temperatures and we initially ran 5 ps of NVT followed by \SIrange{50}{200}{\pico\second} of NPT dynamics, depending on the temperature, with longer simulations in the vicinity of the phase transformation. See SI for further details. A supercell with 1080 atoms, constructed as an $3\times 3 \times 2$ expansion of the 60-atom primitive cell of the LT phase, was used in the simulations. Tests were performed with a larger 3600-atom supercell, see Fig.~\ref{supp-fig:size_conv} in the SI, and we found negligible differences in the extracted lattice parameters compared to the 1080-atom cell. 

Example i-PI input files are openly available~\cite{gitrepo}.
\section{\label{sec:results}Results and Discussion}

\subsection{Reference Models}
We start by performing a series of MD simulations at temperatures ranging from \SIrange{100}{400}{\kelvin} using the MACE-MPA-0 pre-trained foundation model. The resulting average lattice $a$, $b$ and $c$ lattice parameters are shown in Fig.~\ref{fig:mace-mpa}, alongside the experimental results from Ref.~\cite{ammonium_sulfate2015}. The results clearly show that the foundation model is unable to reproduce the qualitatively experimentally observed behavior. Indeed, while it seems to predict a phase transformation at a similar temperature as the experimental one, this transformation is clearly to the wrong state, with both the $a$ and $b$ lattice parameters changing in the opposite direction across the transformation  as compared to the experiments.

\begin{figure}
    \includegraphics[width=\linewidth]{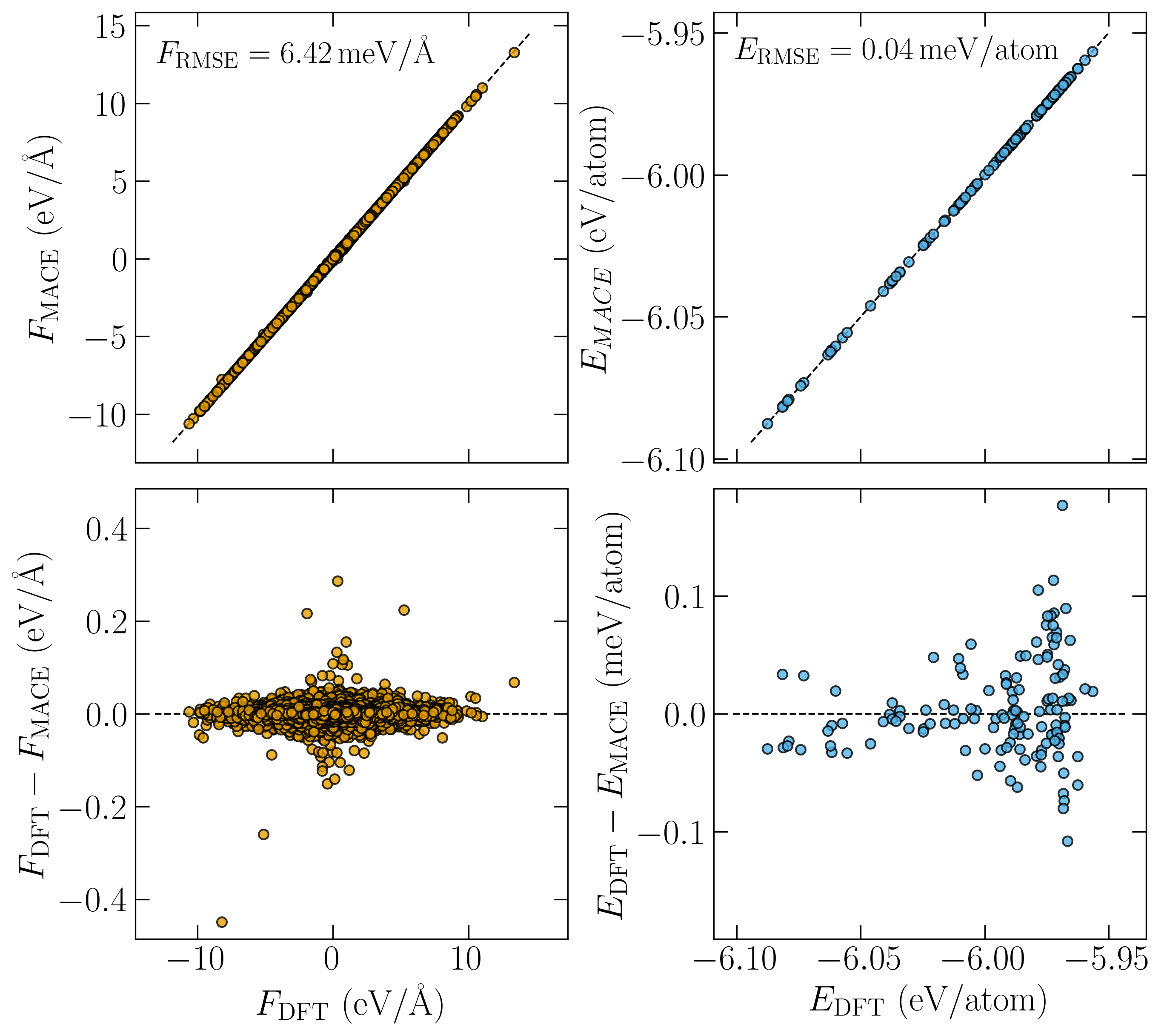}
    \caption{\justifying Parity plots (top) and corresponding prediction errors (bottom) for the reference model trained from scratch on the 360-atom training set evaluated on a hold-out test set, shown for forces (left) and energies (right).}
    \label{fig:parity}
\end{figure}

Next, we consider our reference models trained from scratch on the 360-atom and 60-atom training sets described in Section~\ref{sec:datasets}. Both models show high accuracy on the hold-out test set, as summarized in Table~\ref{tab:rmse_table}, and parity plots and prediction errors for the model trained on the 360-atom datasets are shown in Fig.~\ref{fig:parity}.

\begin{figure*}
    \includegraphics[width=\linewidth]{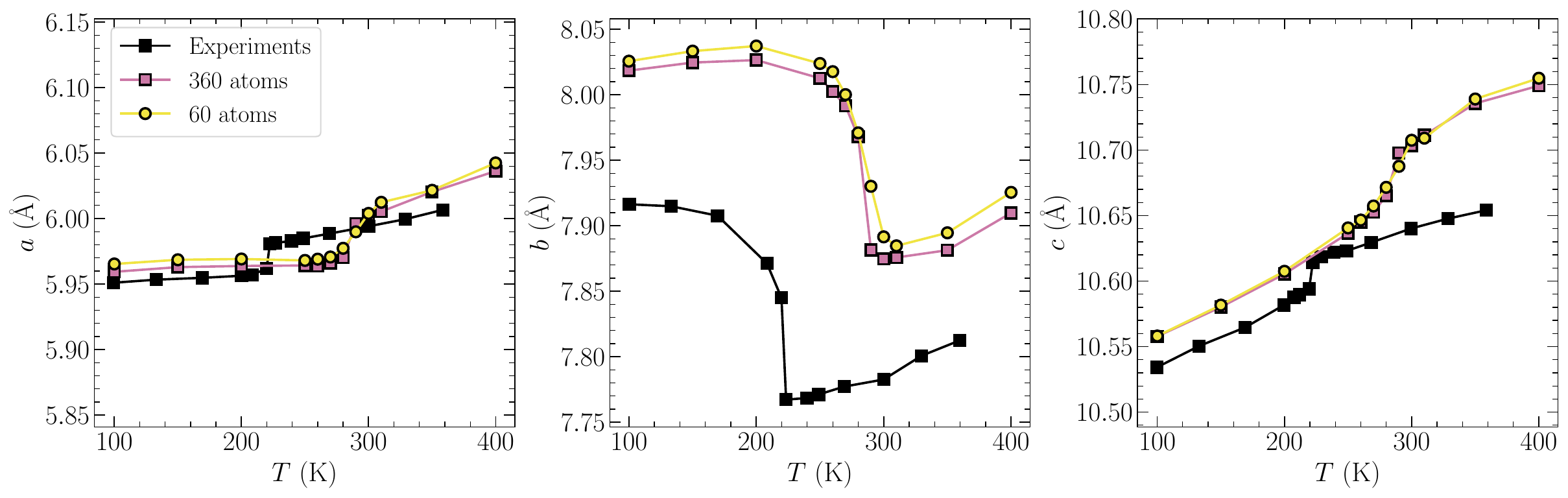}
    \caption{\justifying Temperature dependence of the lattice parameters $a$, $b$, and $c$ in ammonium sulfate, obtained from MD simulations using MACE models trained from scratch on the 360-atom and 60-atom training set, respectively. Experimental values of the lattice parameters from Ref.~\cite{ammonium_sulfate2015}.}
    \label{fig:360_vs_60ats}
\end{figure*}

Fig.~\ref{fig:360_vs_60ats} shows the lattice parameters obtained from MD simulations using these models, together with the experimental results. Unlike the MACE-MPA-0 foundation model, these models reproduce the qualitative behavior of the phase transformation and also predict the lattice parameters relatively well at most temperatures. The most prominent discrepancies are observed for $b$, which is overestimated by $\sim\SI{0.1}{\angstrom}$ and $c$ in the high temperature phase, also overestimated by a similar amount. Furthermore, we notice that the phase transformation temperature is shifted by $\sim\SI{60}{\kelvin}$ compared to the experimental transformation point. 

These discrepancies are likely due to deficiencies in the underlying DFT description provided by the $\mathrm{r^2SCAN}$ functional, rather than the model itself. Corroborating evidence for this is found in the fact that our simulations reveal only a negligible difference between the models trained on 360-atom and 60-atom training set. This despite these models being trained on substantially distinct datasets differing both by cell size and dataset aquisition methodology.   Furthermore, both reference models yield similarly low errors on the same 360-atom hold-out test set. This indicates that 60-atom structures can be used in the training of our models with negligible loss of accuracy. Motivated by this, we employ a strategy of fine-tuning models to 60-atom DFT-data, but evaluate their performance on our hold-out test-set of 360-atom DFT data.

\begin{table}[t]
\centering
\caption{\justifying \label{tab:rmse_table}
Force, energy, and stress RMSEs for the reference models trained from scratch on the 360- and 60-atom datasets, and a selection of the models obtained from fine-tuning on the generated subsets. The size of the subsets are specified inside "()".}
\begin{tabular}{|l|c|c|c|}
\hline
Model & Force RMSE & Energy RMSE & Stress RMSE \\
      & (meV/\AA) & (meV/atom) & (kbar) \\
\hline
360 Ref. model & 6.42 & 0.04 & 0.37 \\
60 Ref. model  & 10.46 & 0.088 & 0.31 \\
Scratch(100)   & 40.54 & 0.56 & 0.68 \\
Scratch(10)    & 128.42 & 7.41 & 3.28 \\
NFT(100)       & 20.99 & 0.11 & 0.42 \\
NFT(10)        & 37.12 & 0.56 & 3.02 \\
MHFT(100)      & 22.15 & 0.34 & 0.85 \\
MHFT(10)       & 35.32 & 0.33 & 1.30 \\
\hline
\end{tabular}
\end{table}

\subsection{Fine-tuned models}
With the goal of investigating the data efficiency of different fine-tuning strategies, we constructed a series of MACE models using both the NFT and MHFT strategies and compared these to models trained from scratch with the same amount of data. 

\begin{figure}
    \includegraphics[width=\linewidth]{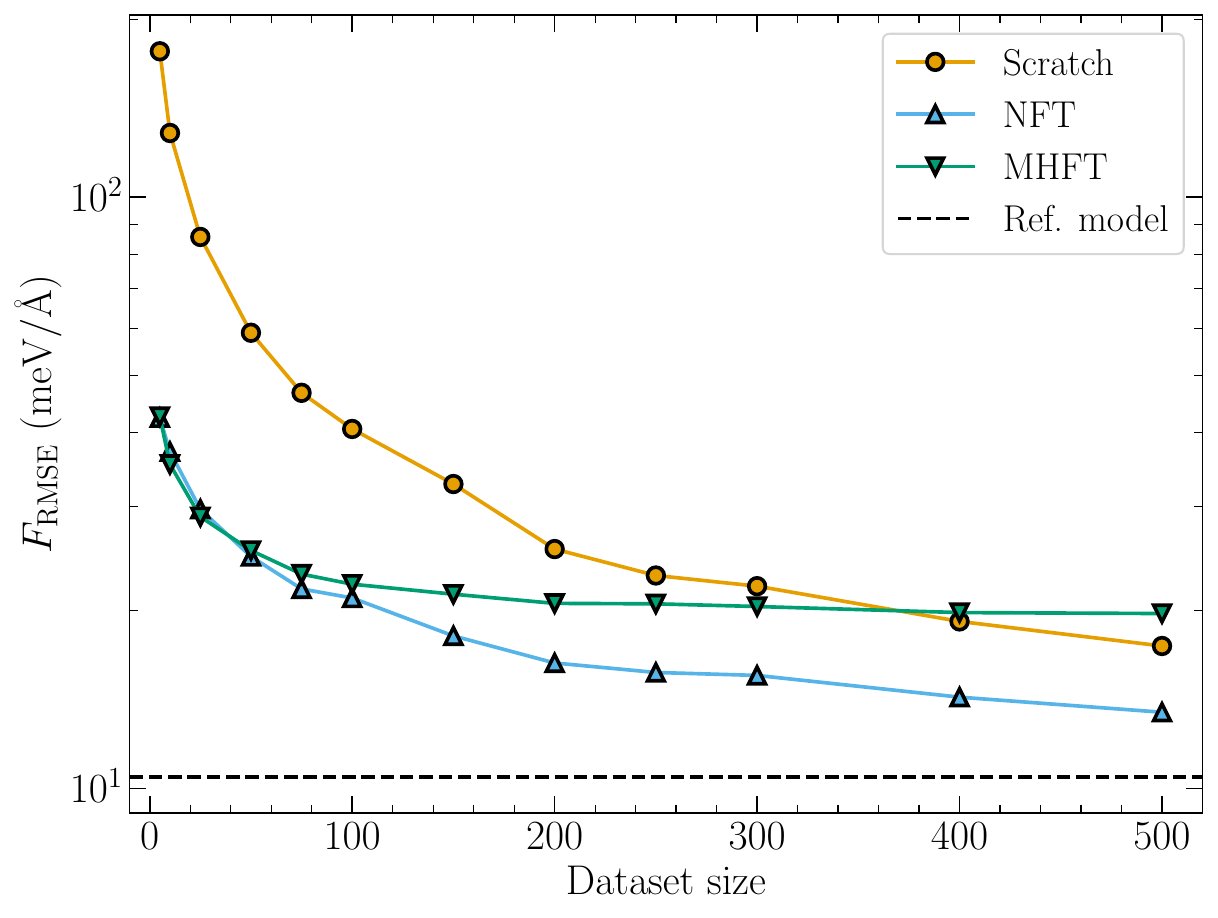}
    \caption{\justifying Comparison of force RMSE as a function of dataset size for three training strategies: training from scratch, naive fine-tuning (NFT), and multihead replay fine-tuning (MHFT), evaluated on the hold-out test set of 360-atom structures. The reference model ("Ref.\ model") is the model trained from scratch on the full 60-atom training set.}
    \label{fig:rmse}
\end{figure}

Fig.~\ref{fig:rmse} shows force RMSEs as a function of dataset size for the different training strategies, and RMSEs on energies, forces and stresses for a select subset of models are shown in Table \ref{tab:rmse_table}. Note that the test set is comprised of 360-atom structures, while these models were all fine-tuned on 60-atom structures. 

In the low-data regime,  the fine-tuning approaches perform significantly better than training from scratch. Indeed, the fine-tuned models show remarkably high accuracy in the very low-data regime, see Table \ref{tab:rmse_table}. 
and the two fine-tuning approaches behave similarly for these small datasets.

As the dataset size increases, the scratch trained models predictably become progressively more accurate, approaching the reference model trained on the full 2403 structure dataset.  

Interestingly, the NFT and MHFT models behave differently for dataset sizes beyond $\sim$ 100. While the errors of the NFT keep decreasing up to the largest dataset sizes we have used, the MHFT models reach a floor in accuracy, with essentially no change after $\sim$ 200 samples. This is likely a result of the MHFT protocol, which constrains the shared MACE descriptors to simultaneously reproduce the pre-training and fine-tuning data, and thereby limits the extent to which the model can specialize to the fine-tuning set. The NFT models do not carry this constraint and can thus adapt more fully to the target data. 

A clear result is that NFT results in smaller model errors for larger datasets. It is important to note however, that this comes at a cost of decreased out-of-domain generalizability, as we show in Section \ref{sec:outofdomain}.

We also find, as seen in Fig.~\ref{supp-fig:random_sampling_rmse} in the SI, very similar model errors when randomly sampling the data from the full training set, compared to the DIRECT sampled ones shown in Fig.~\ref{fig:rmse}.

We note that these quantitative results are likely somewhat sensitive to the detailed settings of different training protocols, in particular the number of training epochs. Nevertheless, the qualitative trends identified above are likely to hold over a range of reasonable training parameters. 

\begin{figure*}
    \includegraphics[width=\textwidth]{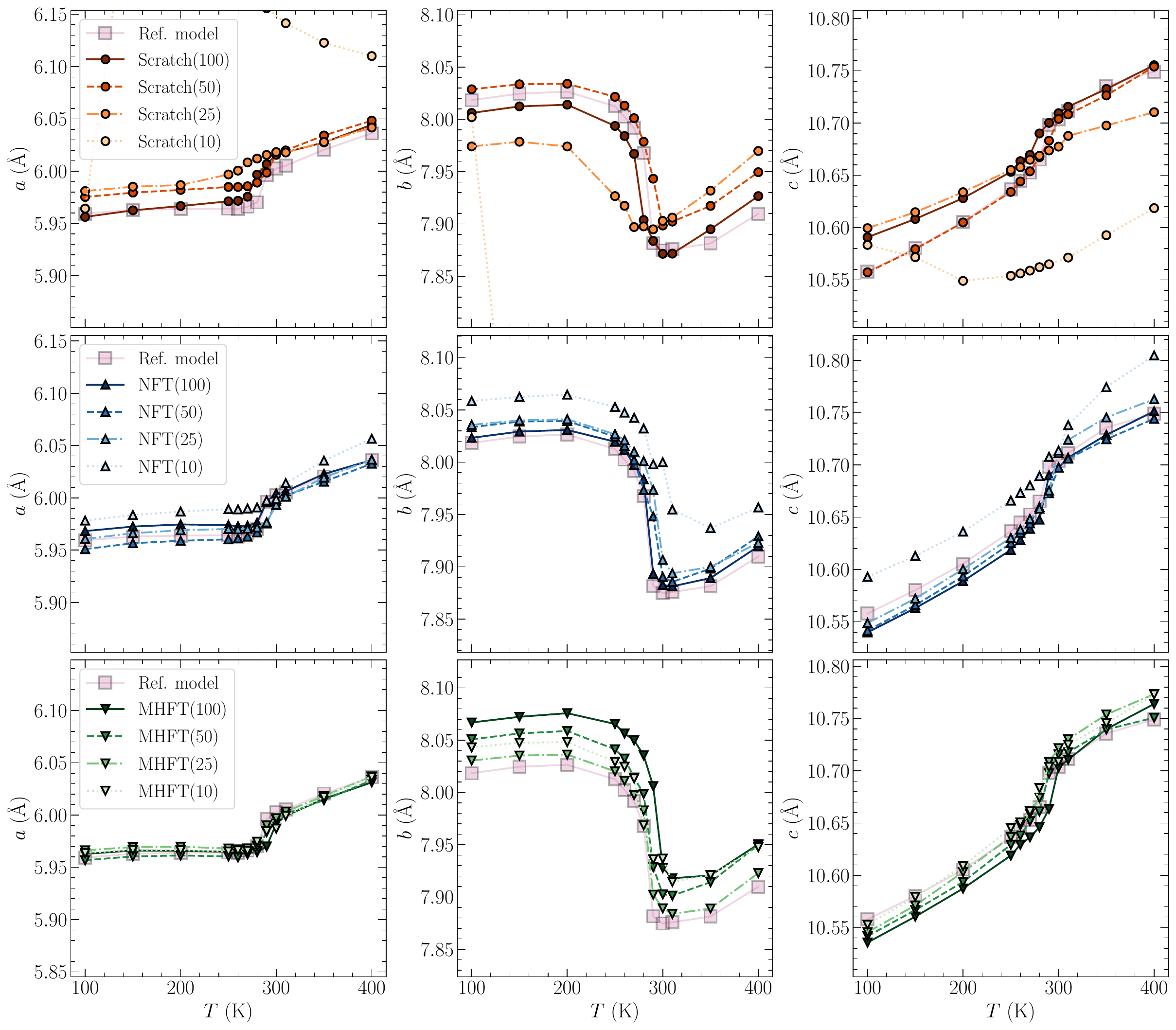}
    \caption{\justifying Temperature dependence of the lattice parameters $a$, $b$, and $c$ of ammonium sulfate, obtained from MD simulations using models trained on datasets of varying sizes ranging from 10 to 100 60-atom DFT configurations, and different training strategies. The top-, middle- and bottom rows show from scratch trained modles NFT models and MHFT models, respectively. Models are denoted by the training strategy, followed by the training set size in parenthesis. The reference model ("Ref.\ model") is the model trained from scratch on the full 360-atom training set.}
    \label{fig:nft-vs-mhft}
\end{figure*}

We next selected the models that were fine-tuned to 100, 50, 25, and 10 60-atom DFT configurations, to perform MD simulations. As a comparison, we also performed MD simulations using the corresponding models trained from scratch. The results from these MD simulations are compiled in Fig.~\ref{fig:nft-vs-mhft}. The fine-tuned models are able to capture the qualitative behavior of the phase transformation, where all lattice parameters agree relatively well with the reference model. In the case of the scratch-trained models, the performance is severely worsened as the dataset size is decreased. It results in relatively good agreement for 100 and 50, but breaks down for 25 and particularly 10 DFT training configurations. Surprisingly, there is a discrepancy between the Scratch(100) and the reference model in the low temperatures of the $c$ parameter, which is not seen in the Scratch(50) model. Moreover, noteworthy is that fine-tuning a foundation model rather than training from scratch yields significantly better accuracy and more stable models in the low-data regime.

The NFT(100), NFT(50) and NFT(25) models yield lattice parameters that are almost identical to those of the reference model, particularly notable in the $b$ parameter while comparing with MHFT. It may indicate that NFT converges toward training from scratch as the size of the training set increases, see Fig.~\ref{fig:nft-vs-mhft}. This is quite conceivable as these two training protocols are inherently very similar and virtually only differ in terms of what model weights to initialize the training from. Thus, it is reasonable that the prior knowledge of the foundation model becomes less important with larger training sets in NFT and increasingly adapts to the fine-tuning data. In contrast, MHFT by construction retains the prior knowledge of the foundation model through the replay head.

Furthermore, a notable shift is observed in the lattice parameters, particularly $b$ and $c$, for NFT(10) compared to NFT(25), NFT(50) and NFT(100), see Fig~\ref{fig:nft-vs-mhft}. This trend is not seen in the MHFT models, although there is a discrepancy between these models in lattice parameter $b$. In addition, a slight instability in the vicinity of the phase transformation is seen for the models fine-tuned on 10 DFT configurations in lattice parameter $b$, which is slightly more pronounced in NFT than in MHFT. However, it should be noted that these differences are negligible and that all fine-tuned models perform quite well with respect to reproducing the phase transformation. We should also emphasize that both fine-tuning protocols yield models trained on only 10 DFT configurations that are still able to reproduce the qualitative behavior of the phase transformation with good agreement, contrary to scratch training. This is rather significant considering the poor performance of the foundation model by itself, highlighting the advantages of fine-tuning.

\begin{figure*}
    \includegraphics[width=\textwidth]{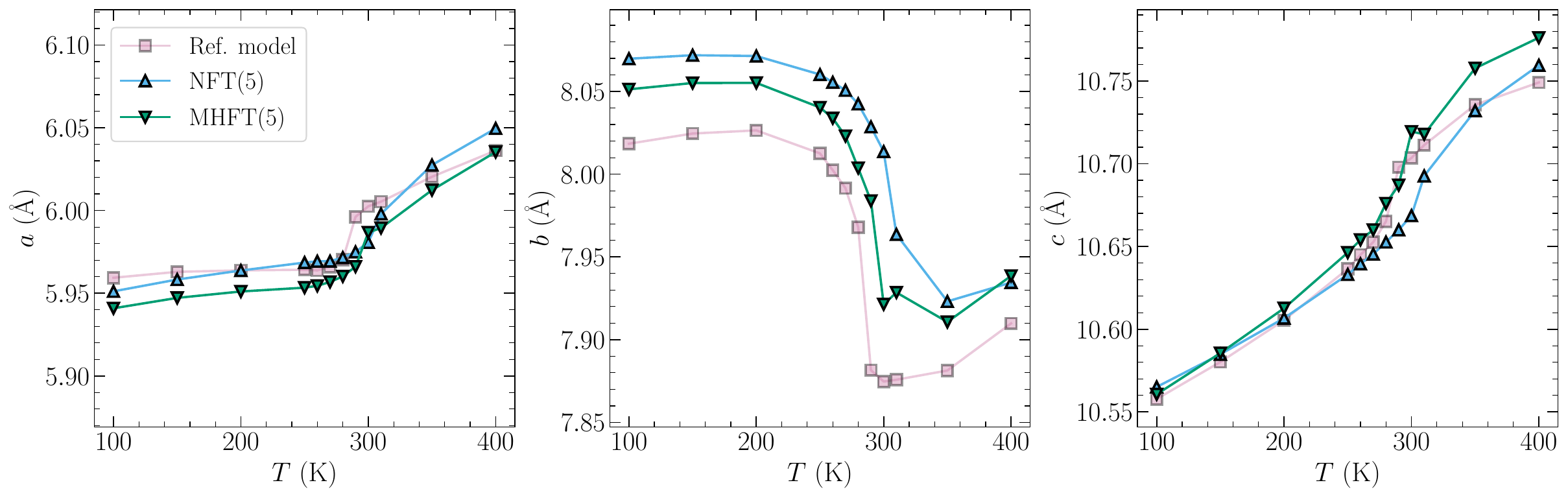}
    \caption{\justifying Temperature dependence of the lattice parameters $a$, $b$, and $c$ of ammonium sulfate, obtained from MD simulations using models fine-tuned to 5 DFT configurations. The reference model ("Ref.\ model") is the model trained from scratch on the full 360-atom training set.}
    \label{fig:ft5}
\end{figure*}
We ran additional MD simulations for the models fine-tuned to 5 DFT training samples since NFT(10) and MHFT(10) in Fig.~\ref{fig:nft-vs-mhft} reproduced the qualitative behavior of the phase transformation with good precision. These simulations are shown in Fig.~\ref{fig:ft5} and also in this case we see that both models are able to reproduce the phase transformation. It is remarkable that 5 reference structures is sufficient considering that the MACE-MPA-0 model by itself was wide of the mark. 

By comparing to Fig.~\ref{fig:nft-vs-mhft}, NFT(5) is slightly shifted up in lattice parameter $b$ compared to NFT(10), similar to what is observed in NFT(10) compared to NFT(25). The HT phase also seems to stabilize slightly later for NFT(5). This may be well within the error margin, or it could potentially be a trend that may slightly indicate that MHFT is somewhat more reliable for very small datasets.

\subsection{Hybrid-DFT models and dispersion corrections}
\begin{figure*}
    \includegraphics[width=\textwidth]{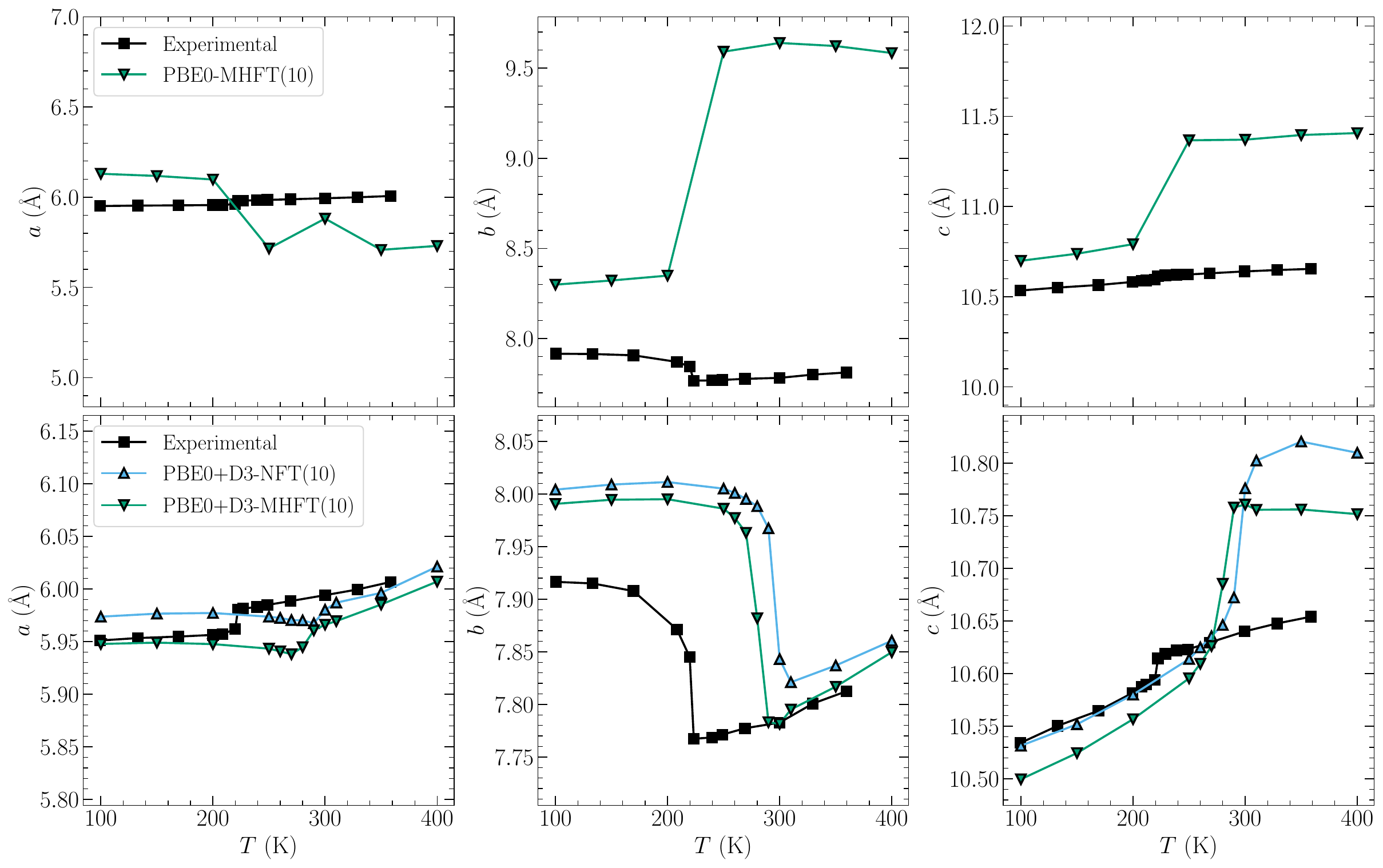}
    \caption{\justifying Temperature dependence of the lattice parameters $a$, $b$, and $c$ in ammonium sulfate, obtained from MD simulations using models fine-tuned to 10 (PBE0) DFT configurations, with (bottom) and without (top) D3-correction.}
    \label{fig:pbe0}
\end{figure*}
Motivated by the high data efficiency of fine-tuning, we fine-tuned the MACE-MPA-0 model to 10 DFT calculations based on the hybrid PBE0 functional, which we used to perform further MD simulations. Initially, DFT calculations with PBE0 were performed without any dispersion correction, but this was shown to be inadequate to describe the phase transformation as seen in Fig.~\ref{fig:pbe0}. However, adding a D3 dispersion correction~\cite{D3} on top of PBE0 appears to at least partially resolve these issues.

This demonstrates that some type of dispersion-correction is necessary to achieve an acceptable description of the phases in ammonium sulfate. This is a reasonable conclusion as the phase transformation is predominantly driven by a reorganization of the hydrogen bonds~\cite{ammonium_sulfate2021}.

Interestingly, explicit dispersion correction was not needed for the $\mathrm{r^2SCAN}$-trained models to describe the phase transformation. The likely reason is that meta-GGAs, and $\mathrm{r^2SCAN}$ in particular, incorporate some intermediate range dispersion interactions in its exchange description~\cite{r2scan-dispersion, vdw-corrections-in-functionals}. This could also offer a partial explanation to the failure of the MACE-MPA-0 foundation model in describing the transformation, since MACE-MPA-0 was trained on dispersion-free PBE-level data.

Further examination of Fig.~\ref{fig:pbe0} and comparison with Fig.~\ref{fig:nft-vs-mhft} does not reveal any clear improvements when using PBE0+D3 over $\mathrm{r^2SCAN}$ in this case. PBE0+D3 provides a somewhat better description of lattice parameter $b$, but overestimating lattice parameter $c$ by a larger margin in the HT phase. However, the main point here is not necessarily to compare and evaluate different functionals, but rather to emphasize that the high data efficiency of fine-tuning enables the use of more sophisticated functionals when generating training data. Consequently, as shown here, it is relatively easy to train MACE-models with the accuracy of hybrid functionals due to the low amount of training data required when fine-tuning a foundation model.

\subsection{\label{sec:outofdomain} Out-of-domain samples}
Finally, we investigate to what extent our fine-tuned model retains knowledge from the foundation models beyond the targeted datasets to which they were trained. 
As mentioned above, the MHFT strategy can alleviate some of the forgetting expected to occur in the NFT approach. Towards this end we compiled a subset of the MPtrj~\cite{MPtrj} dataset containing the same constituent elements as ammonium sulfate (H, S, N, and O),  but excluding datapoints with the precise ammonium sulphate stoichiometry. This makes sure that this data is not present in our fine-tuning dataset, but make up a (small) subset of the original MACE-MPA-0 training set and thus also of the replay set in the MHFT models. We then evaluated these structures using r$^2$SCAN DFT calculations consistent with our fine-tuning training sets.

We note that this setup evaluates two, partly separate, capabilities of our fine-tuned models: 1) to what extent the fine-tuned models retain foundation model knowledge on chemistries and structures beyond the dataset, and 2) the fine-tuned models' ability to translate this knowledge from PBE labels to r$^2$SCAN labels. 

\begin{figure}
    \centering
    \includegraphics[width=\linewidth]{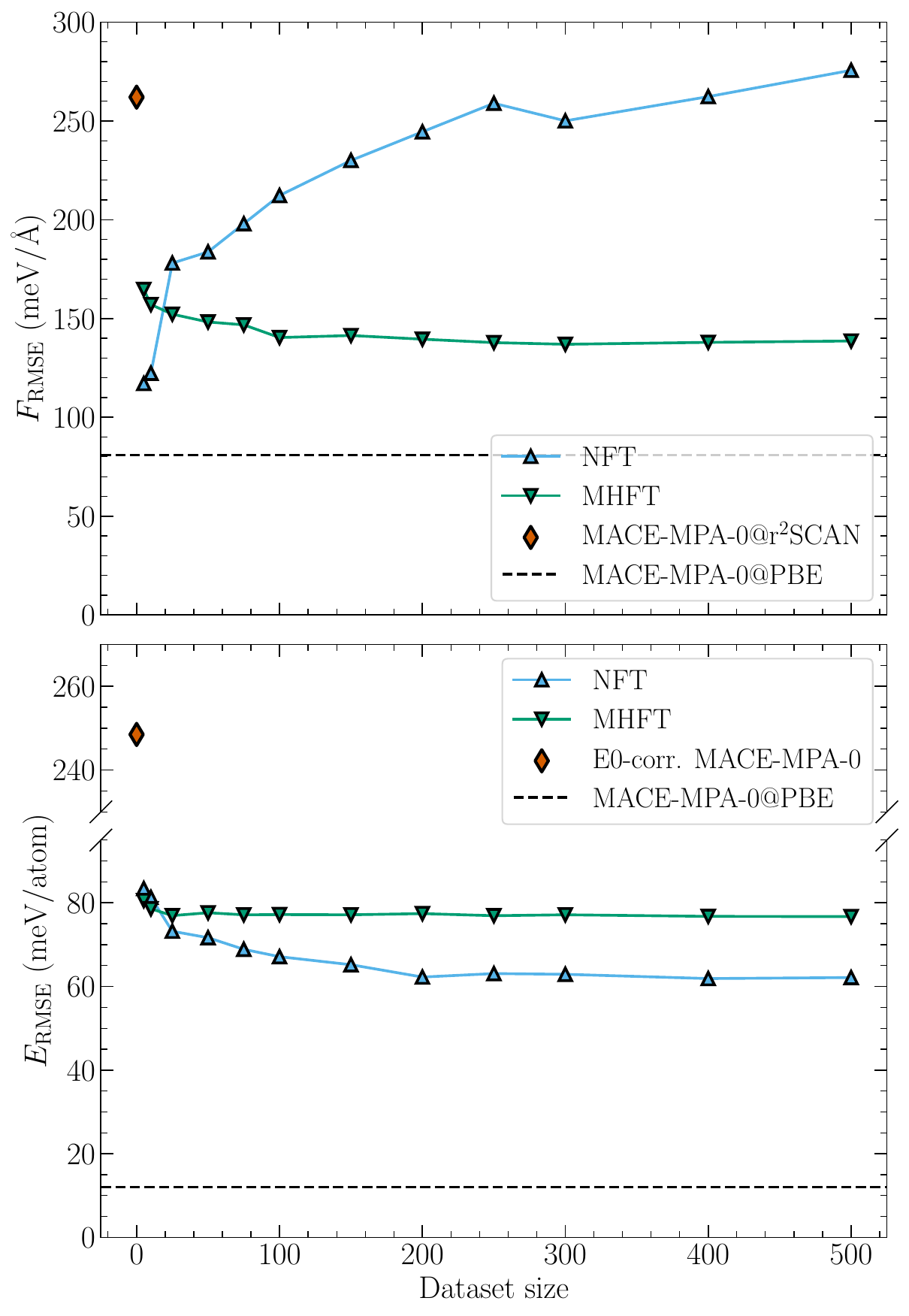}
    \caption{\justifying Force and energy RMSE as a function of dataset size for NFT and MHFT models evaluated on an out-of-domain test set. See text for details on how this test set was constructed.}
    \label{fig:hsno_rmse}
\end{figure}

Fig.~\ref{fig:hsno_rmse} shows the RMSE in forces (top) and energies (bottom) of the NFT and MHFT models for varying training set sizes. We also include two reference datapoints: MACE-MPA-0@PBE corresponding to MACE-MPA-0 evaluated on the PBE-labeled dataset, and MACE-MPA-0@r$^2$SCAN in which the model is evaluated on the r$^2$SCAN DFT labels, where the energy evaluations are adjusted to correct the isolated atomic energies (the MACE E$_0$ parameters) to the r$^2$SCAN level. In accordance with developer recommendations~\cite{mace_multihead_docs}, spin-polarized E$_0$s are used. This can be abstractly viewed as fine-tuning to an empty dataset, i.e.\ just correcting the E$_0$s.

%~\footnote{We note that, in accordance to developer recommendations~\cite{mace_multihead_docs}, we use spin-polarized E$_0$s.}.

The energy errors of the MHFT and NFT models show similar behavior as functions of dataset sizes, both decreasing towards a floor. This decrease is likely primarily a result of the increased ability of the fine-tuned models to adjust the PBE to r$^2$SCAN energy shifts as the amount of r$^2$SCAN they have seen increases. The MHFT models naturally bottoms out at a higher floor than the NFT models due to the replay head, which partially opposes this energy readjustment.

The force RMSE behaviour is, instead, opposite for the two fine-tuning protocols. For NFT, the force RMSE increases with dataset size, while the MHFT the errors initially decrease and then, similarily to the energies, reach a floor. This is a clear sign of forgetting in the NFT models, and further a clear indication that the MHFT scheme mitigates this issue. 

The behaviour can be understood as follows, the drop in $F_{RMSE}$ from the MACE-MPA-0@r$^2$SCAN to the low-data NFT and MHFT models is likely a result of the PBE to r$^2$SCAN readjustment described above. As the dataset size increases, the NFT models are increasingly losing the prior knowledge of the foundation models as they become more specialized to the ammonium-sulfate-specific fine-tuning domain. For the MHFT models, this decrease in out-of-domain performance with an increased dataset size is absent, again an effect of the replay head.

This behavoiur clearly suggest that forgetting effects are clearly present in the NFT approach whereas MHFT appears to be more robust to this effect and that MHFT strategy generalizes better to unseen materials. Nevertheless, the NFT models clearly generalize better than the scratch trained models, indeed our reference model, trained on the full 360-atom dataset produces errors of 792 meV/atom and 1493 meV/Å in energies and forces, respectively.

We also note that these results are presumably sensitive to the number of training epochs used in the fine-tuning protocol. Indeed, increasing the number of epochs is likely to cause a decrease in performance on these out-of-domain samples, particularly for the NFT approach. In addition, the relative weighting between force and energy in the loss functions of the respective training protocols may influence the results.

\section{\label{sec:conclusion} Conclusions}
To summarize, we have investigated how the training-set sizes and training strategy affect a MACE models ability to reproduce the phase transformation of barocaloric ammonium sulfate. Our main finding is that fine-tuning the MACE-MPA-0 foundation model to as little as 5 to 10 DFT (r$^2$SCAN) configurations of 60 atoms produces models able to qualitatively reproduce the phase transformation, despite MACE-MPA-0 itself failing. Models trained from scratch instead require on the order of 50 configurations for a reasonable description.

We find, for our specific test case, that the  fine-tuning strategy is of secondary importance. Indeed, the discrepancy between NFT and MHFT are found to be small; however, the results do provide some indication that MHFT is the better option in the low-data regime. On the other hand, we do find clear signs that MHFT models generalize better to out-of-domain data, consistent with its replay-based protection against catastrophic forgetting,  while clear signs of forgetting are present for the NFT models. A broader exploration of multiple materials could potentially reveal a more pronounced distinction between NFT and MHFT.

The demonstrated data efficiency of foundation model fine-tuning allows for construction of models using more computationally expensive, beyond semi-local DFT, training data. We demonstrate this by constructing models at the PBE0 hybrid-DFT level. We find, however, that these models are unable to reproduce the phase transformation, failing in a similar way as MACE-MPA-0. When training on dispersion corrected data, PBE0+D3, we recover models able to describe the transformation. This demonstrates the need for some description of dispersion to describe the phase-behavior in ammonium sulfate and r$^2$SCAN based models likely work due to some intermediate-range description in its exchange term.

Taken together, our results suggest a potential practical route towards computational screening of larger number of potential barocaloric materials: a foundation model fine-tuned to a few carefully selected reference configurations is sufficient to capture the transformation. We note, however, that the phase transformation in ammonium sulfate is resolvable from MD trajectories on the order of 100 ps, while other materials may require significantly longer times, and/or the aid of enhanced sampling techniques to resolve the transformations in MD accessible timescales. 

\section{Acknowledgments}

Dr.\ F.\ Trybel is acknowledged for useful discussions. We acknowledge financial support from the Swedish Research Council (VR) grant No. 2024-05888, stiftelsen ÅForsk grant No. 24-775 and the Swedish Government Strategic Research Areas in Materials Science on Functional Materials at Link\"{o}ping University (Faculty Grant SFO-Mat-LiU no. 2009 00971).
The computations were enabled by resources provided by the Swedish National Infrastructure for Computing (SNIC), partially funded by the Swedish Research Council through grant agreement no. 2018-05973.

\section{Data Availability}
The data that support the findings of this article are openly available at~\cite{gitrepo}, including datasets, trained models, and example input files for training and simulations.

\bibliography{main} % Produces the bibliography via BibTeX.

\end{document}